\documentclass{iopart}
\usepackage{iopams}
\usepackage{latexsym,amssymb}
\usepackage{graphicx}
\usepackage{amstext}
\usepackage{algorithmic}
\usepackage{algorithm}
\include{epsf}

\def\(({\left(}
\def\)){\right)}                       
\def\[[{\left[}
\def\]]{\right]}

\newcommand{\be}{\begin{equation}}
\newcommand{\ee}{\end{equation}}
\newcommand{\bea}{\begin{eqnarray}}
\newcommand{\eea}{\end{eqnarray}}

\begin{document}

\title{Belief propagation for graph partitioning}
\author{Petr \v{S}ulc$^{1,2,3}$, Lenka Zdeborov\'{a}$^1$}

\address{$^1$ Theoretical Division and Center for Nonlinear Studies, 
Los Alamos National Laboratory,
Los Alamos, NM 87545, USA}

\address{$^2$ New Mexico Consortium, Los Alamos, NM
  87544, USA}

 \address{$^3$ Faculty of Nuclear Sciences and Physical Engineering, 
 Czech Technical University,  B\v{r}ehov\'a 7,  CZ - 115 19 Prague, 
 Czech Republic}

\ead{sulcpetr@gmail.com, lenka.zdeborova@gmail.com}

\date{\today}

\begin{abstract}
We study the belief propagation algorithm for the graph bi-partitioning problem, i.e. the ground state of the ferromagnetic Ising model at a fixed magnetization. Application of a message passing scheme to a model with a fixed global parameter is not banal and we show that the magnetization can in fact be fixed in a local way within the belief propagation equations. Our method provides the full phase diagram of the bi-partitioning problem on random graphs, as well as an efficient heuristic solver that we anticipate to be useful in a wide range of application of the partitioning problem. 
\end{abstract}



\section{Introduction}

Graph partitioning problem was one of the first optimization problems treated with methods of statistical mechanics of disordered 
systems \cite{KirkpatrickGelatt83,FuAnderson86}. Since then other applications of the theory of spin glasses in hard 
optimization and constraint satisfaction problems attracted a lot of interest and many remarkable results were obtained. As anticipated 
in the early works \cite{FuAnderson86,MezardParisi85}, understanding of the energy landscape and the phase transitions in the space 
of solutions leads to understanding of algorithmic hardness of the problems  \cite{MezardParisi02,KrzakalaMontanari06}, and even more 
remarkably it leads to a development of a new class of heuristic algorithmic techniques \cite{MezardParisi02}. Nowadays, the cavity 
method \cite{MezardParisi01} serves as a state of art technique for understanding random optimization problems, and its application on 
a given instances of the problem is a base for a class of one of the most promising heuristic solvers, known as message passing 
algorithms in computer science. 

Despite all this activity in the field, nor the phase diagram neither a message passing algorithm for partitioning a graph into two groups of a given size has been worked out. The main reason that makes the graph partitioning a tricky problem to treat is the existence of a global constraint that fixes the size of the two groups. The aim of this article is to fill this gap, and give the phase diagram of the 
graph bi-partitioning on sparse random graphs and associated belief propagation algorithm.     

\subsection{Partitioning problem: Setting and applications}

A graph $G(V,E)$ is given by the set of vertices $V$ and edges $E$. If an element $(i,j)$ belongs to the set of edges we say that 
vertices $i$ and $j$ are connected. The graph bi-partitioning problem consists of dividing vertices of the graph into two disjoint 
sets of a given size, so as to minimize the number of connections between vertices from different groups. The problem is known to be 
NP-complete \cite{GareyJohnson79}, and hence there is a good reason to believe that no exact polynomial algorithm exists.

The graph partitioning problem is encountered in many important applications. To give few examples: In an electric circuit design one needs to know on which board to place the different components to minimize the number of links between different boards \cite{KarypisAggarwal99}. In parallel computing one has to partition data and tasks among several processors in order to  minimize the communication between them \cite{Pothen97}. Partitioning is also closely related to data clustering and community detection \cite{GirvanNewman06}. The list could continue for long, and it is hence crucial to develop efficient heuristic algorithms that give good solutions to the problem. 

A large volume of literature on heuristic methods for graph partitioning exists. One of the early fundamental works in the field is \cite{KernighanLin70}, its running time is, however, $O(N^2)$ so it is no longer used in practice. Simulated annealing techniques can be used, see e.g. \cite{BanavarSherrington87,SchreiberMartin99}. A local search based methods such as the extremal optimization of \cite{BoettcherPercus01} were suggested. There is a whole class of spectral partitioning methods that use the eigenvectors of the Laplacian of the connectivity graph, see e.g. \cite{HagenKahng92}. However, the current state of art method for partitioning, that is used in most practical applications, is based on the multi-level programming: The nodes are grouped into super-nodes and the super-nodes grouped again, at the end the system size is very small and the problem is solved exactly and the grouping of nodes is then unwrapped. The multi-level programs use elements from many other approaches, see \cite{KarypisKumar99} for an excellent review.

We do not anticipate that belief propagation developed in this paper, will be by itself competitive with the highly tuned implementations of the multi-level methods. However, we do anticipate that it can be used as a component of these implementations. For example, in the multi-level algorithms one needs to estimate the probability that two nodes can be grouped in the same super-node --- this is exactly what belief propagation is designed to compute very fast and efficiently.

The graph bi-partitioning problem is equivalent to finding the ground state of the Ising model with fixed magnetization. The energy in the Ising model is given by the following Hamiltonian: 
\begin{equation}
    H  = - \sum_{(ij)\in E} S_i S_j \, ,   \label{Ham}
\end{equation}
where $S_i$ is the Ising spin (either $+1$ of $-1$) on the $i$-th vertex of the graph. The magnetization $m$, $-1 \le m \le 1$, is given by
\begin{equation}
 \frac{1}{N} \sum_i S_i = m\, ,
\label{eq_m}
\end{equation}
where $N$ is the number of vertices. Therefore, the problem of finding a configuration of spins that minimizes \eref{Ham_h}
while demanding magnetization $m$ to be fixed is equivalent to dividing vertices into two groups of size $N(1+m)/2$ and $N(1-m)/2$ such that the number of links between them is minimal. For $m=0$, the graph is divided into two groups of equal size, i.e. the graph bisection.
The cost of a graph partitioning at a given magnetization, that we call $b(m)$, is given as the number of edges between different groups divided by the total number of vertices. The relation between $b(m)$ and ground state energy $E(m)$ of the Ising model at magnetization fixed to $m$ is
\begin{equation}
\label{BEM}
 b(m)= \frac{E(m) + M}{2N}\, ,  
\end{equation}
where $N$ is the number of nodes, and $M$ the number of edges.

\subsection{Previous results on bi-partitioning random graphs}

In graph theory estimating the asymptotic size of the bisection width in random regular graphs, i.e. graphs of a fixed degree chosen uniformly at random from all the possible ones, is a classical question. Many upper and lower bounds were derived.   
The currently best known upper and lower bounds on bisection width in random regular graphs are by \cite{Bollobas88,KostochkaMelnikov92,MonienPreis01,DiazDo03,DiazSerna07} and we summarize their numerical values in Table \ref{Table1} and Fig.~\ref{fig6}.

For Erd\H{o}s-R\'enyi random graphs with $N\to \infty$ vertices and mean degree $\alpha$ (every edge is present with probability $\alpha/(N-1)$), the size of the largest component is $g N + o(N)$, where $g$ satisfies the following equation:
\begin{equation}
   g= 1 - e^{-\alpha g}.
\end{equation}
In order to divide the graph into two parts of size $N(1+m)/2$ and $N(1-m/2)$ such that the number of edges between the two is zero, the size of the largest component $g$ must be at maximum $(1+m)/2$. That is possible for average degree $\alpha<\alpha_s$ where
\begin{equation}
\label{alp}
  \alpha_s = -\frac{2}{1+m} \log{\frac{1-m}{2}}.
\end{equation}
For $\alpha>\alpha_s$ an extensive number of edges needs to be cut in the minimal bipartition. The value $\alpha_s$ is hence in a sense the satisfiability threshold for graph partitioning of Erd\H{o}s-R\'enyi random graphs. This is further discussed in \cite{PercusIstrate08}, where the authors also obtain an interesting upper bound on the bisection width ($m=0$).

In statistical physics many articles addressed the random graph bi-partitioning problem, see e.g. \cite{FuAnderson86,BanavarSherrington87,SherringtonWong87,WongSherrington87,MezardParisi87c,WongSherrington88,Liao88,GoldschmidtDominicis90}, but as far as we can tell they address only cases where (A) the magnetization is fixed to zero, (B) the fluctuations in the degree of the random graph are negligible, i.e. the graphs are either dense or regular. The computational techniques used in the above mentioned papers do not generalize to the non-zero magnetization case nor to graphs with fluctuating degree, as e.g. to the Erd\H{o}s-R\'enyi random graphs. We will give a more detailed explanation of why the techniques do not generalize in section \ref{reg_results}. This also justifies novelty of the approach developed in this article.

\subsection{Contribution of this article}

If the ground state energy of the ferromagnetic Ising model (\ref{Ham}) was a convex function of the magnetization $m$ then an external magnetic field (playing the role of the chemical potential from the grand-canonical ensemble) could be used to compute $E(m)$ with a standard cavity method \cite{MezardParisi03}. However, random graphs are mean field topologies and the energy at fixed magnetization $E(m)$ does not have to be and in this case is not a convex function, similarly as in the fully connected Curie-Weiss model. The problem of imposing the value of the magnetization is hence more challenging.

A method to explore the non-convex parts of thermodynamical potentials within the Bethe-Peierls (Belief Propagation) approximation was suggested in \cite{DiMontanari04}, and used later e.g. in \cite{MoraMezard06,KrzakalaRicci09}. The main idea is to introduce an uniform external magnetic field (or chemical potential) and adjust its value after every update of the local cavity fields. We use this method for partitioning graphs, and we argue that it (or its generalization to the replica symmetry breaking scheme) is asymptotically exact on sparse random graphs. 

The main practical contribution of this article is the belief propagation algorithm for graph partitioning problem that we believe to be of use in the various applications of the problem. We study the behavior and performance of the algorithm on random graphs but we anticipate it will be meaningful and useful for other families of graphs, complex networks for example. 

We also compute the phase diagram of (\ref{Ham}) at fixed magnetization. In \cite{PercusIstrate08} it was argued that in the Erd\H{o}s-R\'enyi graphs at zero magnetization the glassy transition happens at some average degree strictly larger than the satisfiability threshold, $\alpha_c>\alpha_s$, we indeed confirm this conjecture, we compute $\alpha_c$ and several other quantities of interest. 

An interesting side remark, discussed in section \ref{reg_results}, concerns the case treated in the previous works: the regular random graphs at zero magnetization. There the average properties of the graph bi-partitioning are equivalent to those of the spin glass problem. We argue why this equivalence does not generalize to non-zero magnetization or non-regular graphs. More detailed discussion about the equivalence can be found in \cite{BoettcherZdeborova09}. 

\section{Cavity method at fixed magnetization}

As we explained in the introduction, the graph partitioning is equivalent to the ferromagnetic Ising model at fixed magnetization $m^*$. The magnetization will be fixed via an external magnetic field $h$ which appears in the Hamiltonian as
\be
     H_h  = - \sum_{(ij)\in E} S_i S_j - h \sum_i S_i \, .  \label{Ham_h}
\ee
The ground state energy density of \eref{Ham} and \eref{Ham_h} are related via the Legendre transformation $e(h) =   e(m) -   hm$, so that the parameter $h$ has to be chosen such that 
\be
       \frac{\partial e(h)}{\partial h}\Big|_{h^*} = -m^* \, . \label{f_der}
\ee
If $e(h)$ is the ground energy density of \eref{Ham_h} with field $h$ corresponding to magnetization $m$, the corresponding partition cost (\ref{BEM}) of the graph is 
\begin{equation}
\label{BE}
 b= \frac{e(h)+ h m + \frac{\alpha}{2}}{2} \, ,
\end{equation}
where $\alpha$ is the mean degree of the graph.

\subsection{Belief propagation equations}

The Bethe-Peierls approximation, or the Belief-Propagation equations, aims to describe the Boltzmann measure of (\ref{Ham_h})
\be
        \mu(\{S_i\}) = \frac{e^{-\beta H_h(\{S_i\})}}{Z}\, , 
\ee
where $\beta$ is the inverse temperature. The graph partitioning problem corresponds to $\beta\to \infty$. In this section we summarize the well known belief propagation equations for this problem. For a detailed derivation see \cite{YedidiaFreeman03,MezardMontanari07}.  

In the most standard form of belief propagation equations \cite{YedidiaFreeman03} one introduces $\psi_{S_i}^{i\to j}$ to be the probability that variable $i$ takes value $S_i$ given the interaction on $(ij)$ is absent. On a tree (cycle free) graph then 
\be
   \psi_{S_i}^{i\to j}= \frac{1}{Z^{i\to j}} e^{\beta h S_i}   \prod_{k\in \partial i \setminus j}  \Big(\sum_{S_k}  e^{\beta S_i S_k} \psi_{S_k}^{k\to i} \Big) \, ,   \label{BP_prob} 
\ee
where $Z^{i\to j}$ is normalization ensuring $\psi_{+1}^{i\to j}+\psi_{-1}^{i\to j}=1$. After a fixed point of eqs.~(\ref{BP_prob}) is found the Bethe free energy (or the log-partition function) is given as \cite{YedidiaFreeman03}
\be
    -\beta F(h)= \log{Z} = \sum_i \log{Z^i} - \sum_{(ij)} \log{Z^{ij}}\, , \label{free_en}
\ee
where 
\bea
   Z^i &=& \sum_{S_i} e^{\beta h S_i}   \prod_{k\in \partial i}  \Big(\sum_{S_k}  e^{\beta S_i S_k} \psi_{S_k}^{k\to i} \Big) \, , \label{zi}\\
   Z^{ij} &=& \sum_{S_i,S_j}  e^{\beta S_i S_j} \psi_{S_i}^{i\to j} \psi_{S_j}^{j\to i}\, . \label{zij}
\eea
At a given value of the external magnetic field $h$ the average magnetization is computed as $m = -[\partial F(h) /\partial h] /N$, using (\ref{free_en}) one gets
\be
      m = \frac{1}{N} \sum_i \frac{ \sum_{S_i} S_i \, e^{\beta h S_i}   \prod_{k\in \partial i}  \Big(\sum_{S_k}  e^{\beta S_i S_k} \psi_{S_k}^{k\to i} \Big)   }{ \sum_{S_i} e^{\beta h S_i}   \prod_{k\in \partial i}  \Big(\sum_{S_k}  e^{\beta S_i S_k} \psi_{S_k}^{k\to i} \Big)   }\, . \label{mag_T}
\ee

In order to write the zero temperature limit, $\beta \to \infty$, of the above equations we introduce more suitable messages (usually called cavity fields) $h^{i\to j}$
\be
        e^{2\beta h^{i\to j}} \equiv \frac{\psi_{+1}^{i\to j}}{\psi_{-1}^{i\to j}}  \label{cavity_f}
\ee
One then obtains equations equivalent to the replica symmetric equations in \cite{MezardParisi03}. The self-consistent equations for messages (\ref{BP_prob}) become
\be
 \hspace{-1cm}  h^{i\to j} = h +  \sum_{k\in \partial i \setminus j} \left[ \max{(1+h^{k\to i},0)} - \max{(h^{k\to i},1)} \right] \equiv {\cal F}(\{h^{k\to i}\}) \, . \label{BP_0T}
\ee
Note that the term in the sum is $-1$ for $h^{k\to i}\le -1$, $+1$ for $h^{k\to i}\ge 1$, and $h^{k\to i}$ for $-1<h^{k\to i}< 1$. The Bethe estimate of the ground state energy is
\be
   E(h)= \sum_i {E^i} - \sum_{(ij)} {E^{ij}}\, , \label{ET0}
\ee
where from (\ref{zi}-\ref{zij}) we obtain
\bea
     { E}^i &=& h + \alpha + 2 \sum_{k\in i  } \max{(0,h^{k\to i})} \nonumber \\ &-& 2\max{[h+\sum_{k\in i }\max{(1+h^{k\to i},0)},\sum_{k\in i  }\max{(h^{k\to i},1)} ]} \label{ei} \\
     { E}^{ij} &=& 1+2\max{(0,h^{i\to j})}+2\max{(0,h^{j\to i})}
\nonumber \\ &-& 2 \max{(1+h^{i\to j}+h^{j\to i},1,h^{j\to i},h^{i\to j})} \, .\label{eij}
\eea
And finally the contribution to the sum $\sum_i$ in the expression for the magnetization (\ref{mag_T}) is in the zero temperature limit equal to $+1$ if 
\be
      h +  \sum_{k\in \partial i} \left[ \max{(1+h^{k\to i },0)} - \max{(h^{k\to i },1)}\right] > 0\, ,
\ee
and $-1$ otherwise (if the two terms are equal the contribution is $0$). For notation let us call this function 
\be
          m = {\cal M}(h,\{h^{i\to j}\})\, .  \label{mag_0}
\ee

\subsection{Population dynamics at fixed magnetization}
In order to calculate the average ground state energy (\ref{ET0}), and thus the partitioning cost $b$, for a given ensemble of random graphs one implements the population dynamics method \cite{MezardParisi01,MezardMontanari07}. 

In the standard population dynamics one would update eqs.~(\ref{BP_0T}) with a given value of external magnetic field $h$ till convergence or till  maximum number of iterations and then one would compute the ground energy and the corresponding magnetization. If this is done with the above equations for graph bi-partitioning then the resulting magnetization will always be either $+1$ for $h>0$ or $-1$ for $h<0$. We however want to find the ground state energy at magnetization values $-1<m^*<1$. In order to do that we will not keep the external field $h$ constant. Instead after every iteration of (\ref{BP_0T}) we use the current valued of fields $h^{i\to j}$ and update the value of $h$ in such a way that $m^* = {\cal M}(h_{\rm new},\{h^{i\to j}\})$ where $m^*$ is the desired value of the magnetization. The resulting population dynamics code is sketched in algorithm \ref{alg_pop}.  

\begin{algorithm}
\caption{Population dynamics algorithm for BP on $d$-regular random graphs with fixed magnetization $m^*$}
\label{alg_pop}
\begin{algorithmic} 
 \STATE $h \gets 0$
\STATE Randomly initialize messages $h(i)$, $i= 1,2 \ldots N$ 
\FOR{$j = 1$ to max}
 \FOR{$i = 1$ to $N$}
    \STATE  Randomly select $d$ indices in $k= 1,2 \ldots N$  
    \STATE  Calculate $h(i)$ from $\{h(k)\}$ using eq.~\eref{BP_0T}
 \ENDFOR 
   \STATE  $h_1 \gets h - 1$
    \STATE  $h_2 \gets h + 1$ 
     \WHILE{$|h_1- h_2|<$ criterion}
        \STATE Calculate magnetization $m$ with external field $h$ using eq.~\ref{mag_0}            
         \IF{$m < m^*$}
              \STATE $h_1 \gets h$
          \ENDIF
           \IF{$m > m^*$}
              \STATE $h_2 \gets h$
          \ENDIF
       \STATE   $h \gets  (h_1 + h_2)/2$
     \ENDWHILE
\ENDFOR
\STATE Calculate $E$ using \eref{ET0} (averaged over some number of iterations)  
\RETURN $E$, $h$
\end{algorithmic}
\end{algorithm}

Note that the algorithm \ref{alg_pop} uses bisection method in each iteration in order to fix the magnetization. For $d$-regular graphs, ${\cal M}(h,\{h^{i\to j}\})$ as a function of $h$ for given values of $\{h^{i\to j}\}$ is continuous monotonic function and therefore the algorithm always manages to fix the desired magnetization. 
For general graphs ${\cal M}(h,\{h^{i\to j}\})$ may have less well behaved form and we will discuss that in section \ref{sec_heuristic}.

\subsection{1RSB at fixed magnetization}
\label{sec:1RSB}

As may be anticipated from the relation between graph bisection and the spin glass \cite{FuAnderson86} the belief propagation equations (replica symmetric approach) are not always asymptotically exact for the graph bi-partitioning. Instead in some regions of parameters the problem is glassy and the replica symmetry breaking approach is needed for an exact solution, just like in the Sherrington-Kirkpatrick model \cite{SherringtonKirkpatrick75,Parisi80b}.
The replica symmetry breaking approach for sparse random graphs was developed in \cite{MezardParisi01,MezardParisi03} and is well established. Hence we only point out the difference in the equations that leads to fixing a non-trivial value of the magnetization.  

In order to write the 1RSB equations we follow closely the approach of \cite{MezardParisi03}. We introduce a complexity function $\Sigma(E)$, i.e. number of thermodynamical states at a given energy, and its Legendre transform $\Phi(y)$ also called the replicated potential 
\be
      - y \Phi(y) = - y E + \Sigma(E)\, ,  \quad \frac{\partial y \Phi(y)}{\partial y} = E \label{1RSB_pot} \, .
\ee
Every thermodynamical state has a corresponding value of the cavity field $h^{i\to j}$ and according to \cite{MezardParisi03} the self-consistent equation for the distribution of cavity fields over states is
\be
\hspace{-1.5cm} P^{i\to j}(h^{i\to j}) = \frac{1}{{\cal Z}^{i\to j}} \int \prod_{k\in \partial i \setminus j} {\rm d}P^{k\to i}(h^{k\to i}) e^{-y E^{i\to j}} \delta[h^{i\to j} - {\cal F}(\{h^{k\to i}\})]  \, ,
\label{1RSB}
\ee
where ${\cal F}(\{h^{k\to i}\})$ is defined by eq.~(\ref{BP_0T}).
The reweighting factor is defined by $E^{i\to j} = - \lim_{\beta\to \infty}\frac{1}{\beta} \log{Z^{i\to j}}$ where $Z^{i\to j}$ is the normalization constant in (\ref{BP_prob}) and is given by an equation analogous to (\ref{ei}).
Once a fixed point of (\ref{1RSB}) is found the potential $\Phi(y)$ is computed as follows $\Phi(y) = \sum_i \Phi^i - \sum_{ij} \Phi^{ij}$ with
\be
  e^{-y \Phi^i} = \int_{POP}  e^{-y { E}^i} \, , \quad \quad
  e^{-y \Phi^{ij}} = \int_{POP}  e^{-y { E}^{ij}} \, , 
\ee
where the notation $\int_{POP} = \int \prod_{k\in \partial i \setminus j} {\rm d}P^{k\to i}(h^{k\to i})$ and the energy contributions are given by (\ref{ei}-\ref{eij}).
The energy of the system is then computed according to (\ref{1RSB_pot}) as $E=\sum {\cal E}^i - \sum_{ij} {\cal E}^{ij}$ with
\be
   {\cal E}^i = \frac{\int_{POP} E^i e^{-y { E}^i}   }{\int_{POP}  e^{-y { E}^i} } \, ,\quad \quad 
   {\cal E}^{ij} = \frac{\int_{POP} E^{ij} e^{-y { E}^{ij}}   }{\int_{POP}  e^{-y { E}^{ij}} } \, .
\ee
And the magnetization $m=\sum_i m^i/N$, where
\bea
    m^i &=& - \frac{\partial {\cal E}^i }{\partial h} = - \frac{\int_{POP} \frac{\partial E^i}{\partial h} e^{-y { E}^i}  }{\int_{POP}  e^{-y { E}^i} }  +y \frac{\int_{POP}\frac{\partial E^i}{\partial h} E^i  e^{-y { E}^i} }{\int_{POP}  e^{-y { E}^i} } \nonumber \\ &-& y \frac{\int_{POP} E^i  e^{-y { E}^i} \int_{POP} \frac{\partial E^i}{\partial h}  e^{-y { E}^i}}{(\int_{POP}  e^{-y { E}^i} )^2} \, . \label{mag_1RSB}
\eea
Note that $ \frac{\partial E^i}{\partial h} = \pm 1$ depending on the sign in eq.~(\ref{mag_0}).

Again the only difference between the usual 1RSB and 1RSB at fixed magnetization is that after every iteration the external magnetic field is chosen a new value such that magnetization computed from (\ref{mag_1RSB}) is equal to the desired value $m^*$.

Solving the 1RSB equations is often tedious and to obtain the phase diagram it is often sufficient to investigate the convergence of the belief propagation iterations. This is equivalent to analyzing the local stability of the replica symmetric solution towards replica symmetry breaking, as done originally by de Almeida and Thouless \cite{AlmeidaThouless78}. Within the population dynamics we use the noise propagation method (for a derivation see appendix C of \cite{Zdeborova09}). In the population dynamics algorithm together with cavity fields $h^{i\to j}$, one keeps track of the quantity
\begin{equation}
 v^{i\to j} = \sum_{k \in \partial i \setminus j}  \frac{ \partial  h^{i\to j} }{   \partial h^{k\to i}}   v^{k\to i}
\end{equation}
after every sweep of BP iteration we normalize the values $v^{i\to j}$ by $\lambda$ in such a way that $\sum (v^{i\to j}/\lambda)^2 = 1$. Parameter $\lambda$ then plays a role of a certain Lyapunov exponent and the belief propagation does not converge if and only if on average $\lambda>1$. We have found that BP never converges on regular graphs for any value of magnetization $-1<m^*<1$. Nevertheless, the value of the energy calculated with BP gives a good lower bound on the actual energy of the model \cite{FranzLeone01}. For the Erd\H{o}s-R\'enyi graphs with given magnetization, we found a phase transition from a replica symmetric region where BP is asymptotically exact to a glassy region where RSB solution would be need to obtain the asymptotically exact value of the ground state energy (this phase transition is shown in Fig.~\ref{fig_4}).

\section{BP as a heuristic solver}
\label{sec_heuristic}

Equations for the belief propagation derived in the previous section can be used on a given graph as we sketch in algorithm \ref{alg_fin}. 

\begin{algorithm}
\caption{BP algorithm for partitioning of a given graph}
\label{alg_fin}
\begin{algorithmic} 
  \STATE $\forall i,j$ Initialize messages $h^{ i \to j}$ and field $h$ randomly
  \STATE iter  $\gets 0$
 \REPEAT 
   \FORALL{  $i \in V $}
    	\STATE convergence $\gets 0 $
        \STATE { \rm local\_field[i] }  $\gets \sum_{k\in \partial i} \left[ \max{(1+h^{k\to i},0)} - \max{(h^{k\to i},1)} \right] $ 
        \STATE  $h_{\rm new}^{i \to j} \gets   h + $ local\_field(i) - $h^{j \to i}$ 
        \STATE convergence $\gets$ convergence $+ \left| h_{\rm new}^{i \to j} -  h^{i \to j}  \right|$ 
        \STATE $h_{\rm new}^{i \to j} \gets$    memory$*h^{j \to i}$ $+$ (1 $-$  memory)$* h_{\rm new}^{i \to j}$  
    \ENDFOR
    \STATE sort({\rm local\_field})
    \STATE $h$ = - \rm{ local\_field[ $N(1 - m) / 2$] }
   \STATE iter $\gets$ iter $+ 1$
 \UNTIL { convergence $< \epsilon$ OR {\rm iter} $>$ {\rm maximum iterations} }
 \STATE compute $E$ using equation \eref{ET0}
 \RETURN $E$, $h$
\end{algorithmic}
\end{algorithm}

The parameter {\it memory}, which we set to $0.7$ in our simulations, is introduced in order to prevent messages from oscillating. If the algorithm does not converge after a given maximum number of iterations, it is terminated. However, even if the algorithm does not converge, the calculated $E$ still provides a reasonable estimate of the bisection cost that is on average a lower bound of the true average cost. 

In the presented algorithm, we introduced a slightly different method to fix the magnetization by manipulating $h$. In the algorithm 
\ref{alg_fin}, we sort all the local cavity fields and set $h$ so that $N(1 - m) / 2$ of them are negative (or zero) and the rest positive (or zero). It follows from the definition of messages \eref{cavity_f} that the positive value of local cavity field means that spin on this node is more likely to be equal to $1$, negative means that the spin is more likely to be $-1$. If the local cavity field is exactly equal to zero, the spin in a given node 
is unbiased (free). This can be used to actually obtain a graph partition. However a decimation technique,  algorithm \ref{alg_dec}, achieves much better results in particular when many free or almost free spin are present, reported in Figs.~\ref{fig7} and \ref{fig6}. 

\begin{algorithm}
 \caption{Decimation algorithm}
 \label{alg_dec}
\begin{algorithmic}
  \REPEAT
  \STATE Run algorithm \ref{alg_fin} 
  \STATE Choose a vertex $i$ such that local\_field($i$) is the highest (or lowest if this is an even iteration) and fix all outgoing messages from this node to $+ \infty$ (  - $\infty$ for an even iteration). Fix spin in vertex $i$ to $+1$ ($-1$ in even iteration) 
  \UNTIL{ Number of fixed spins to $+1$ or $-1$ reaches the value required to fix desired magnetization $m$ }
\end{algorithmic}
\end{algorithm}

Note that in the present form the decimation solver has running time quadratic in the size of the system. However, linear running time can be achieved without significant loss of performance by decimating a finite fraction of spins after every iteration, as in the survey propagation algorithm of \cite{MezardParisi02}.

\section{Behavior of the method and results}

In this section we discuss the behavior of the belief propagation algorithm of random regular and Erd\H{o}s-R\'enyi random graphs. We, however, anticipate that qualitatively similar behavior as on the  Erd\H{o}s-R\'enyi random graphs will be seen on other graph families.  

\subsection{Phase diagram of Erd\H{o}s-R\'enyi graphs bi-partitioning}

The most interesting to discuss is the behavior of the algorithm on a given graph and the decimation. In particular: Is the function ${\cal M}(h,\{h^{i\to j}\})$ (\ref{mag_0}) continuous in $h$ such that any value of the magnetization can be fixed? Does the external field $h$  converge in the iterations? Does the decimation achieve low energy states? We choose typical Erd\H{o}s-R\'enyi random graphs to illustrate the behavior and answer these questions. 

An Erd\H{o}s-R\'enyi random graph of average degree smaller that one, $\alpha<1$, basically looks like a collection of small disconnected trees. Let us hence first discuss how does the algorithm behave on a tree. On a tree the belief propagation equations (\ref{BP_0T}) have only one possible fixed point for every value of $h$. For $h>0$ all $h^{i\to j}=h+d_i-1$, where $d_i$ is the degree of node $i$ and magnetization $m=1$, for $h<0$ all $h^{i\to j}=h-d_i+1$ and $m=-1$, and for $h=0$ all $h^{i\to j}=0$ and $m=0$. 

When fixing magnetization to some value $-1<m^*<1$ the third fixed point is attractive. But provides a wrong value of magnetization $m=0$, as in this case the function ${\cal M}(h,\{h^{i\to j}\})$ is a step function. The fixed point also does not provide much of a useful information about the cost of splitting a tree on two groups of a given size. So in a sense our algorithm does not behave very well on trees, that is a kind of unusual situation for belief propagation. 
The decimation algorithm, however, works well and is able to obtain reasonably good partitions even on a tree. This is because once a spin is fixed the information propagates and is taken care of correctly.

\begin{figure}[!ht]
\begin{center}
\includegraphics[width=0.49\linewidth]{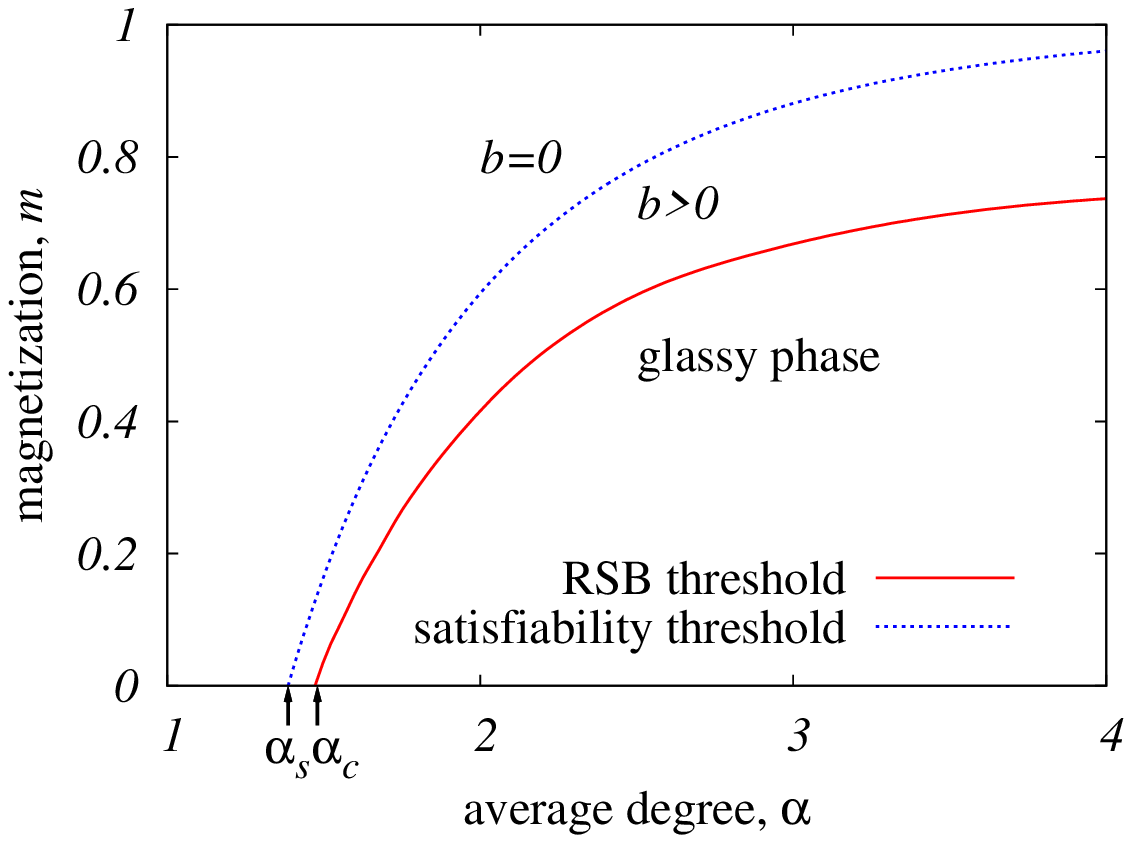}
\includegraphics[width=0.49\linewidth]{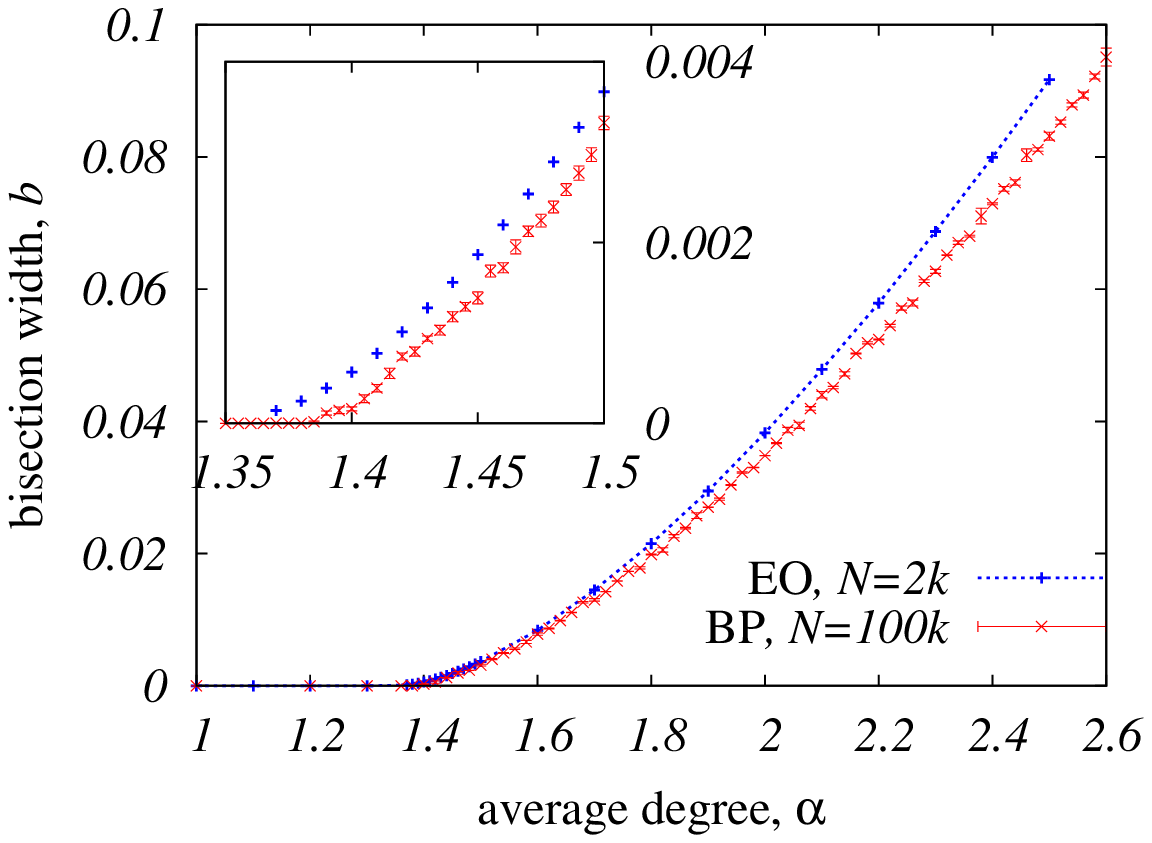}
\end{center}
\caption{Left: The plot shows two phase transitions in partitioning of Erd\H{o}s-R\'enyi random graphs. The satisfiability threshold $\alpha_s$, eq.~(\ref{alp}), above which the giant component have to be cut to bipartition the graph. And the glass transition, $\alpha_c$, at which the belief propagation equations stop to converge and replica symmetry breaking is needed to describe correctly the system. Note that $\alpha_s<\alpha_c(m=0)=1.472$ as anticipated in \cite{PercusIstrate08}. Right: Bisection width $b$ on Erd\H{o}s-R\'enyi graph as a function of the mean connectivity computed by averaging over $2$ graphs of size $N=100000$ with algorithm \ref{alg_fin}. The data are compared with the exact average bisection width $b$ calculated with the extremal optimization heuristics for $N=2000$, data from \cite{PercusIstrate08} \label{fig_4}. As the replica symmetric result provides a lower bound on the energy and the exact ground states on systems of finite sizes are in this case larger that the asymptotic values, the asymptotic value must lie between the two curves. The inset zooms into the phase transition region.}
\end{figure}

But back to Erd\H{o}s-R\'enyi random graphs, for mean connectivities above the percolation threshold but lower than the satisfiability threshold $1<\alpha<\alpha_s$ (given by \eref{alp}, and depicted in Fig.~\ref{fig_4}) one finds that the algorithm \ref{alg_fin} converges to a configuration such that on the giant component of the graph all local 
fields are positive (or negative). Thus all spins on the giant component will be chosen to be +1 (-1). In the rest of the graph (that is all the small components) the local fields as well as the external field $h$ are negative (positive) but very close to zero. In such a case again, the BP algorithm without decimation is not very efficient in actually dividing the components into two properly sized groups. However, the decimation algorithm achieves this task quite well (note that if one spin on a small component is fixed, than all the other vertices orient in the same direction). 

After the satisfiability threshold \eref{alp}, the giant component is bigger than the number of vertices that are in the larger of the two groups, so inevitably one will have neighbors with opposite spins in the ground state. There are two possibilities: (A) BP converges or (B) BP does not converge. If BP does converge, i.e. bellow $\alpha_c$, then it converges to a configuration where the giant component is divided into two groups (positive and negative local fields) and all the other components of the graph are oriented in one direction (the one that has smaller number of vertices on the giant component). In order to fix the proper magnetization on the giant component the external field is nonzero even when the total magnetization $m^*=0$. 

BP does not converge above the replica symmetry breaking threshold $\alpha_c$ depicted in Fig.~\ref{fig_4}. But even in such cases the snapshots of fields are meaningful and the decimation algorithm achieves good energies even when the non-convergence is ignored, as illustrated in section \ref{sec:dec}.

\begin{figure}[!ht]
\begin{center} 
\includegraphics[width=0.49\linewidth]{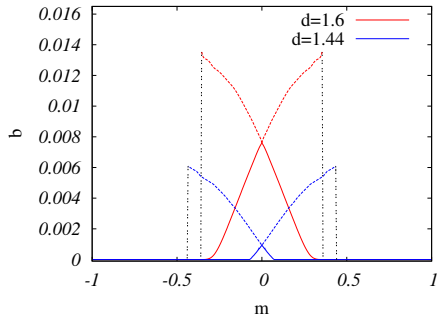}
\end{center}
\caption{The figure shows the partitioning cost $b$ as a function of $m$ for two different Erd\H{o}s-R\'enyi graphs with mean connectivities $1.44$ and $1.6$ (and of sizes $N=100000$). In the simulation, the messages were randomly initialized for $m=-1$ and then $b$ was calculated with 
algorithm \ref{alg_fin}. Magnetization was then slightly increased to $m + \Delta m$ and messages $h^{i\to j}$ were initialized with their values from simulation with previous $m$. The dashed curves correspond to the case when the system orients the spins on the small components in the less favorable way (that is, $+1$ for $m>0$ and vice versa). 
The dashed curves end at a spinodal point where the giant component is divided in half. \label{fig_spinodals}}
\end{figure}

In fact on the Erd\H{o}s-R\'enyi random graphs there is a first order phase transition at zero magnetization. 
At the transition the derivative of the energy with respect to magnetization has a discontinuity. On both sides of the transition a meta-stable state exists with spinodal points at values of magnetization corresponding to the half size of the giant component. This phase transition and lines corresponding to the meta-stable state and the spinodal point are illustrated in the figure \ref{fig_spinodals}. 

How to understand this phase transition: Consider large positive magnetization, in the lowest cost solution the giant component and large part of the small components are positive and a small part of the components are negative. As the magnetization is decreased the small components are all turning negative, and also parts of the giant component turn negative. The external field is negative in that region in order to keep the small components negative. Even after half of the spins become negative the system does not realize that it is less costly to turn everybody, instead if the magnetization is slowly decreased further the belief propagation equations indicate that a larger fraction of the giant component should be negative. As the magnetization is decreased the negative external field becomes closer to zero, at the point the external field flips to positive values the small component turn to positive direction and the system realizes this gives much lower cost. This point corresponds to a spinodal point. Of course this discussion could be repeated by changing the works positive for negative and vice versa. The phase transition, meta-stable state, and spinodal point are illustrated in figure \ref{fig_spinodals}. 

If the magnetization is not changed gradually, depending on the initial conditions the algorithm does converge to one or the other of the branches, more likely to the lower one. This is a nice property as if more stable divisions are present in real network our algorithm might be able to find them (or at least those of them with considerably large basin of attraction).

\subsection{Performance of the BP decimation}
\label{sec:dec}

In this section we illustrate accuracy of the decimation BP solver on random $3$-regular graphs. Regular graphs are in some sense the hardest case for graph bi-partitioning as they look locally alike from every point and no apparent structure can be explored to decide if two nodes should be in the same group or not.

\begin{figure}[!ht]
\begin{center}
\includegraphics[width=0.49\linewidth]{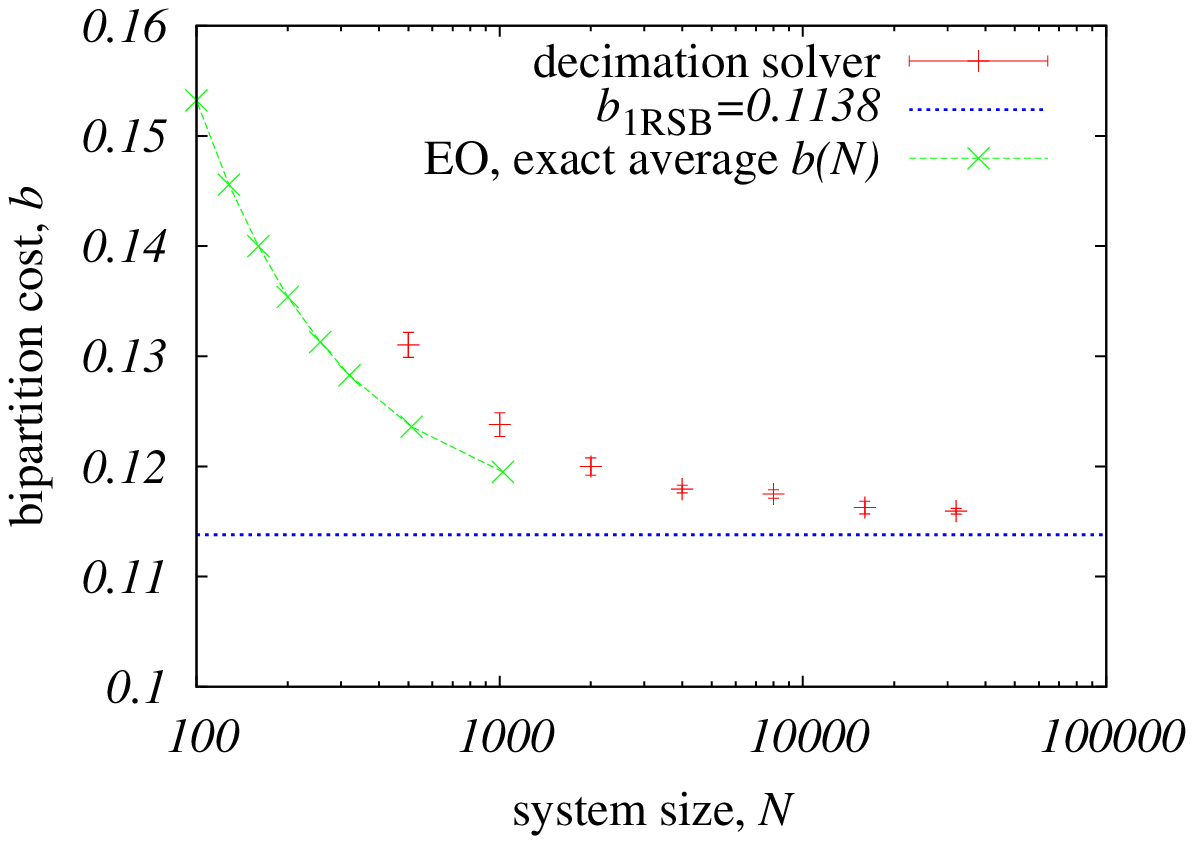}
\includegraphics[width=0.49\linewidth]{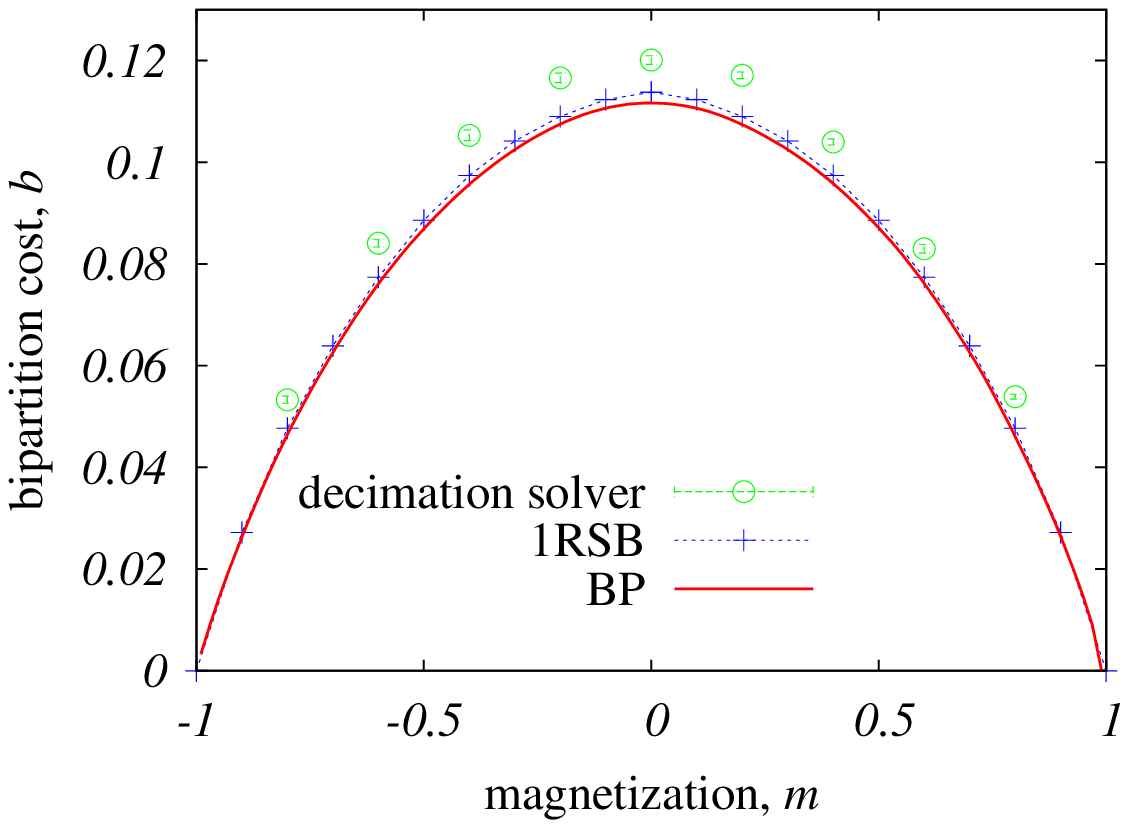} 
\end{center}
\caption{Left: Decimation results for 3-regular random graphs of different sizes, compared to presumably exact average ground state energies as computed from the extremal optimization heuristics by \cite{BoettcherPercus01}. Also shown is the asymptotic cost $b=0.1138$ calculated by 1RSB method. Note that the decimation algorithm is far better than the best known algorithmic bound $b=0.1\overline{6}$ \cite{MonienPreis01}. Right: The plot shows replica symmetric (BP) results, 1RSB results and performance of the decimation algorithm for the partition cost $b$ as a function of the magnetization $m$ for $3$-regular random graphs. The BP population dynamics algorithm was with $N=10000$, 1RSB solutions were obtained from a simulation with $N=30000$. The decimation results were were averaged over $10$ different graphs, each with $N=2000$. \label{fig7}}
\end{figure}

If Fig.~\ref{fig7} we show the average bisection cost achieved by the decimation solvers on graphs of different size. We compare to the asymptotic value of the cost and to the average values obtained from  extremal optimization heuristic of \cite{BoettcherPercus01,BoettcherZdeborova09} that are exact (or at least very close to exact), we see that our decimation solver achieves energies very closed to the ground states. In particular note that the best provable algorithmic bound for 3-regular graph bisection is  $b=0.1\overline{6}$ \cite{MonienPreis01} which is far above what decimation achieves.

In the right part of Fig.~\ref{fig7} we compare the partition cost as a function of the magnetization $m$ as obtained from (a) the population dynamics solving the BP equations, (b) resolution of the 1RSB equations from sec.~\ref{sec:1RSB} under the assumption that for every edge the distribution of fields $P^{i\to j}(h^{i\to j})$ is the same --- this being called the factorized solution in \cite{MezardParisi03}, and (c) decimation solver run on graphs of size $N=2000$.

\subsection{Random regular graphs at zero magnetization}
\label{reg_results}

In this subsection we want to discuss the bisection (zero magnetization) of random regular graphs. This case has been treated in \cite{FuAnderson86,SherringtonWong87,WongSherrington87,MezardParisi87c,WongSherrington88,GoldschmidtDominicis90} using analogy with spin glasses, i.e. Hamiltonian 
\be
       H_{\rm SG}  = - \sum_{(ij)\in E} J_{ij} S_i S_j \, ,   \label{Ham_SG}
\ee
with random $J_{ij}=\pm 1$ has been solved instead of fixing magnetization to zero via an external field. 

Indeed, note that in random regular graph it is more than reasonable to assume that the two groups in graph bisection have exactly the same properties and hence the first order phase transition that we have seen at $m=0$ in the Erd\H{o}s-R\'enyi graph is expected to disappear. Consequently, the slope of the ground state $e(m)$ at $m=0$ is expected to be zero, and hence also the value of external field to which our algorithm converges is zero $h=0$. 

We remind that cavity fields $h^{i\to j}$ can be interpreted as a change in the ground state energy of (\ref{Ham_h}) when link $(ij)$ is removed from the graph. If $h$ is an integer then also all $h^{i\to j}$ have to be integers in the the final solution of the problem. The cavity equations can then be parameterized by fraction of negative, positive and zero cavity fields  $h^{i\to j}$. The only way to achieve zero magnetization is then to set the fraction of negative and positive cavity fields equal. And this leads exactly to the same equations as M\'ezard and Parisi obtained in \cite{MezardParisi03} and justifies the approach of \cite{FuAnderson86,SherringtonWong87,WongSherrington87,MezardParisi87c,WongSherrington88,GoldschmidtDominicis90}. Consequences and generalization of this equivalence will be described in \cite{BoettcherZdeborova09}.

We want to stress that at non-zero magnetization the corresponding  external field $h$ does not take an integer value and hence no straightforward relation to the spin glass problem exists. Also as long as the degree of the graph is not constant there might be a room for a first order phase transition at $m=0$ due to asymmetries between the two groups in the bisection - as illustrated in the Erd\H{o}s-R\'enyi graphs. If the first order phase transition is present that at $m=0$ the external field $h\neq 0$ and hence again no straightforward analogy with the spin glass problem exists. Thus the approach developed in this paper is the only one know that is able to treat non-regular graphs or non-zero values of the magnetization. 
 
\begin{figure}[!ht]
\begin{center}
\includegraphics[width=0.55\linewidth]{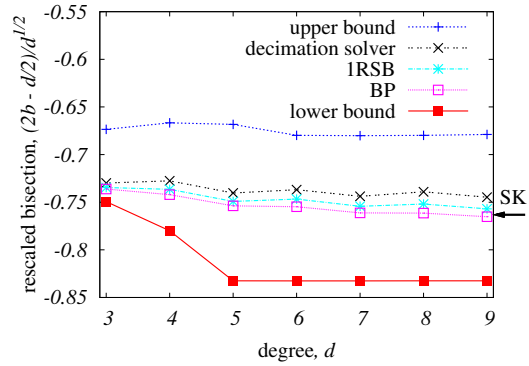}
\end{center}
\caption{We plot data from Table \ref{Table1} rescaled as $(2b-\frac{d}{2})/\sqrt{d}$ as a function of the degree $d$. According to \cite{FuAnderson86} for large $d$ the true values should converge to the ground state energy of the Sherrington-Kirkpatrick model, $E=-0.763219$. \label{fig6}}
\end{figure}

In Table \ref{Table1} and Fig.~\ref{fig6} we summarize the known rigorous bounds for bisection widths in random regular graphs. We also summarize results of belief propagation obtained from our population dynamics, and the results from 1RSB calculation using integer values of the cavity fields. Both the latter are only approximation to the full-step replica symmetry breaking result that would presumably be exact in this case. Finally we compare with performance of our decimation BP solver. In particular Fig.~\ref{fig6} illustrates how accurate the decimation solver is. Not that the true value of the bisection width must lie between the decimation and 1RSB data points.

\begin{table}[!ht]
\begin{center}
\begin{tabular}{|l|l|l|l|l|l|}
\hline
d  &    $b_{\rm low}$   & $b_{\rm up}$   &      $b_{\rm RS}$     &    $b_{\rm 1RSB}$   & $b_{\rm BP dec}$ \\
\hline \hline  
 3  &     0.101  &  0.166$\overline 6$    &       0.1125(2)     &    0.113846    &  0.1180(3) \\ \hline
4   &    0.22   &  0.333$\overline 3$    &        0.2579(2)    &      0.263527  &    0.272(1)  \\ \hline
5   &     0.3192  &  0.5028    &       0.4072(3)     &    0.412398    &  0.422(2)  \\ \hline
6   &     0.4803  &  0.6674    &        0.5756(3)    &   0.585414     & 0.5975(9)   \\ \hline
7   &     0.6486  &  0.8502    &         0.7430(4)   &     0.752171   &   0.766(2)    \\ \hline
8    &    0.8226   & 1.0386    &        0.9232(4)    &     0.936595   &   0.955(2)    \\ \hline
9    &    1.0012   & 1.2317    &        1.1022(4)    &   1.11453      &  1.133(1)    \\ \hline

\end{tabular}
\end{center}
\caption{\label{Table1} This table summarizes the best known lower bound (2nd column, \cite{KostochkaMelnikov92,Bollobas88}, the Bollobas's bound $d/4 - \sqrt{(d\ln 2)}/2$ is the best known for $d\ge 5$) and the upper bound (3rd column, \cite{MonienPreis01,DiazDo03,DiazSerna07}) bounds for random regular graph bisection. In the 4th column we give results for the bisection from the population dynamics for belief propagation, these numbers are identical to the ones obtained in \cite{WongSherrington88} with non-integer cavity fields. The 5th column gives results of the 1RSB calculation with integer fields as developed in \cite{MezardParisi03}. And the final column shown performance of our implementation of the BP decimation algorithm for graphs of size $N=2000$.}
\end{table}

\section{Discussion}

The main practical contribution of this article is the belief propagation algorithm for graph partitioning problem that we anticipate to be useful in the various applications of the partitioning problem. We studied the behavior and performance of the algorithm on random graphs but we anticipate it will be meaningful also for other families of graphs, complex networks in particular. Compared to other partitioning algorithms BP has the advantage that is provides information about probability with which a certain node is in a certain groups. It is also able to see different locally stable divisions of the graph - as illustrated by the first order phase transition in Erd\H{o}s-R\'enyi graphs at zero magnetization. In real world networks the partitioning cost at different values of magnetization $m$ may lead to a non-trivial information about communities in the network and information about their significance. Note also that our approach is straightforwardly generalizable to $k$-partitioning the graph into $k$ groups of a fixed size.

\section*{Acknowledgment}

We thank Stefan Boettcher for sharing with us his data from the extremal optimization algorithm that we used for comparison in Figs.~\ref{fig_4}, \ref{fig7}. We thank Cris Moore for pointing to us the meaning of the first order phase transition at zero magnetization and the existence of the spinodal lines illustrated in Fig.~\ref{fig_spinodals}. We also thank Florent Krzakala, Mark Newman, Allon Percus, and Federico Ricci-Tersenghi for various very useful discussions about this work. P\v{S} acknowledges partial support of New Mexico Consortium via NSF collaborative grant CCF-0829945 on ``Harnessing Statistical Physics for Computing and Communications''

\section*{References}

\bibliographystyle{unsrt}
\bibliography{myentries}

\end{document}